\documentclass[sigconf, screen]{acmart}
\AtBeginDocument{%
  \providecommand\BibTeX{{%
    \normalfont B\kern-0.5em{\scshape i\kern-0.25em b}\kern-0.8em\TeX}}}

\setcopyright{rightsretained}
\copyrightyear{2024}
\acmYear{2024}

\acmConference[arXiv]{arXiv: arXiv Preprint}{2024}{Atlanta, GA}
\acmBooktitle{arXiv: arXiv Preprint, 2024, Atlanta, GA}

\usepackage{enumitem}
\usepackage{xspace}
\usepackage{wrapfig}

\newcommand{\todo}[1]{\textcolor{red}{[[[TODO: #1]]]}}
\newcommand{\tocite}[1]{\textcolor{teal}{[[CITE: #1]]}}
\newcommand{\pratham}[1]{\textcolor{blue}{[#1 -Pratham]}}
\newcommand{\polo}[1]{\textcolor{orange}{[#1 -Polo]}}
\newcommand{\harsha}[1]{\textcolor{purple}{[#1 -Harsha]}}

\newcommand{\tool}[0]{\textsc{ARCollab}\xspace{}}

\newcommand{\mkclean}{
  \renewcommand{\todo}[1]{}
  \renewcommand{\tocite}[1]{}
  \renewcommand{\pratham}[1]{}
  \renewcommand{\polo}[1]{}
  \renewcommand{\harsha}[1]{}
}

\mkclean{}

\begin{document}

\title{\tool{}: Towards Multi-User Interactive Cardiovascular Surgical Planning in Mobile Augmented Reality}

\author{Pratham Darrpan Mehta}
\orcid{0000-0001-6650-8489}
\affiliation{%
  \institution{Georgia Institute of Technology}
  \city{Atlanta}
  \country{USA}}
\email{pratham@gatech.edu}
  
\author{Harsha Dinesh Karanth}
\affiliation{%
  \institution{Georgia Institute of Technology}
  \city{Atlanta}
  \country{USA}}
\email{hkaranth3@gatech.edu}

\author{Haoyang Yang}
\affiliation{%
  \institution{Georgia Institute of Technology}
  \city{Atlanta}
  \country{USA}}
\email{alexanderyang@gatech.edu}

\author{Timothy C. Slesnick}
\affiliation{%
  \institution{Children's Healthcare of Atlanta}
  \city{Atlanta}
  \country{USA}}
\email{SlesnickT@kidsheart.com}

\author{Fawwaz Shaw}
\affiliation{%
  \institution{Children's Healthcare of Atlanta}
  \city{Atlanta}
  \country{USA}}
\email{Fawwaz.Shaw@choa.org}

\author{Duen Horng Chau}
\orcid{0000-0001-9824-3323}
\affiliation{%
  \institution{Georgia Institute of Technology}
  \city{Atlanta}
  \country{USA}}
\email{polo@gatech.edu}

\renewcommand{\shortauthors}{Mehta and Karanth, et al.}

\begin{abstract}
Surgical planning for congenital heart diseases requires a collaborative approach, traditionally involving the 3D-printing of physical heart models for inspection by surgeons and cardiologists. 
Recent advancements in mobile augmented reality (AR) technologies have offered a promising alternative, noted for their ease-of-use and portability. 
Despite this progress, there remains a gap in research exploring the use of multi-user mobile AR environments for facilitating collaborative cardiovascular surgical planning. 
We are developing \tool{}, an iOS AR application designed to allow multiple surgeons and cardiologists to interact with patient-specific 3D heart models in a shared environment. 
\tool{} allows surgeons and cardiologists to import heart models, perform gestures to manipulate the heart, and collaborate with other users without having to produce a physical heart model. 
We are excited by the potential for \tool{} to make long-term real-world impact, thanks to the ubiquity of iOS devices that will allow for \tool{}'s easy distribution, deployment and adoption.
\end{abstract}

\ccsdesc[500]{Human-centered computing~Mixed / augmented reality}

\keywords{Augmented Reality, Mobile Collaboration, Surgical Planning}


\begin{teaserfigure}
  \includegraphics[width=\textwidth]{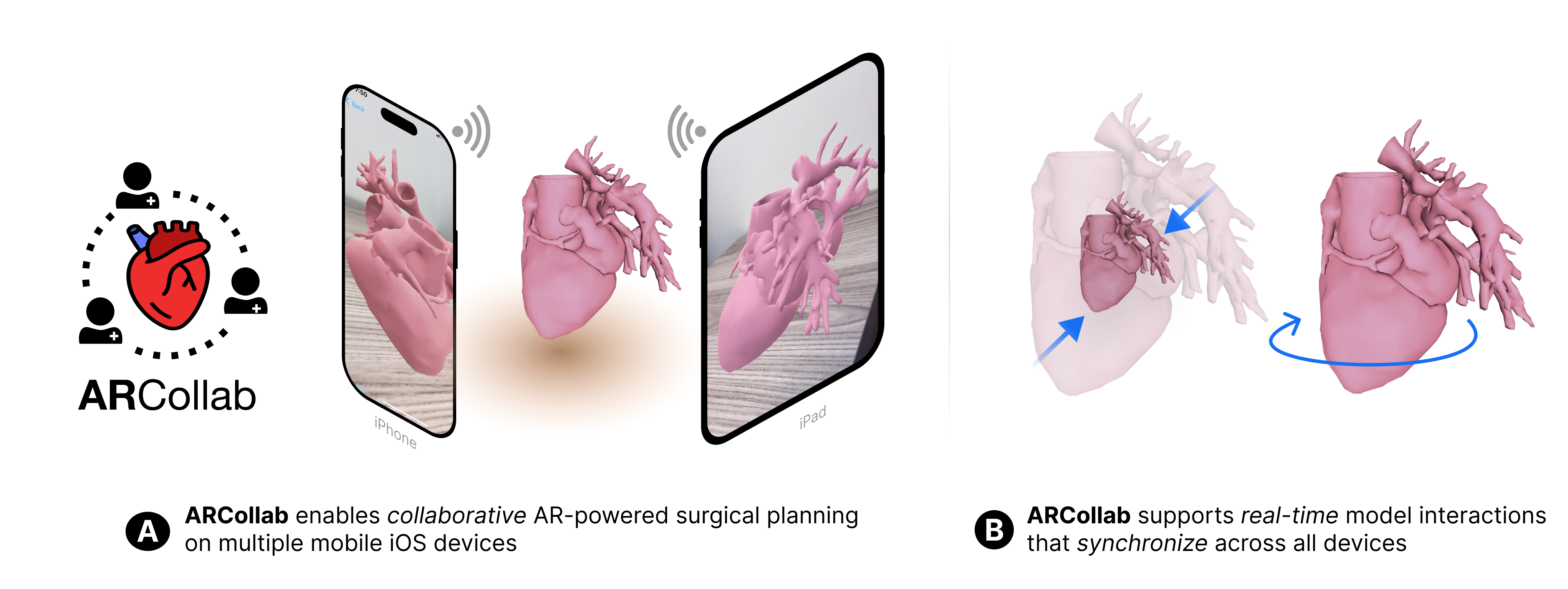}
  \caption{(A) \tool{} is a collaborative cardiovascular surgical planning application in mobile augmented reality. Multiple users can join a shared session and view the a patient's 3D heart model from different perspectives. (B) \tool{} allows surgeons and cardiologists to collaboratively interact with a 3D heart model in real-time. Changes made on a device will update the model's orientation on the other devices in the session.}
  \label{fig:teaser}
\end{teaserfigure}


\maketitle
\pagestyle{plain}

\section{Introduction}

\begin{wrapfigure}{L}{0.18\textwidth}
    \begin{center}
        \includegraphics[width = 0.2\textwidth]{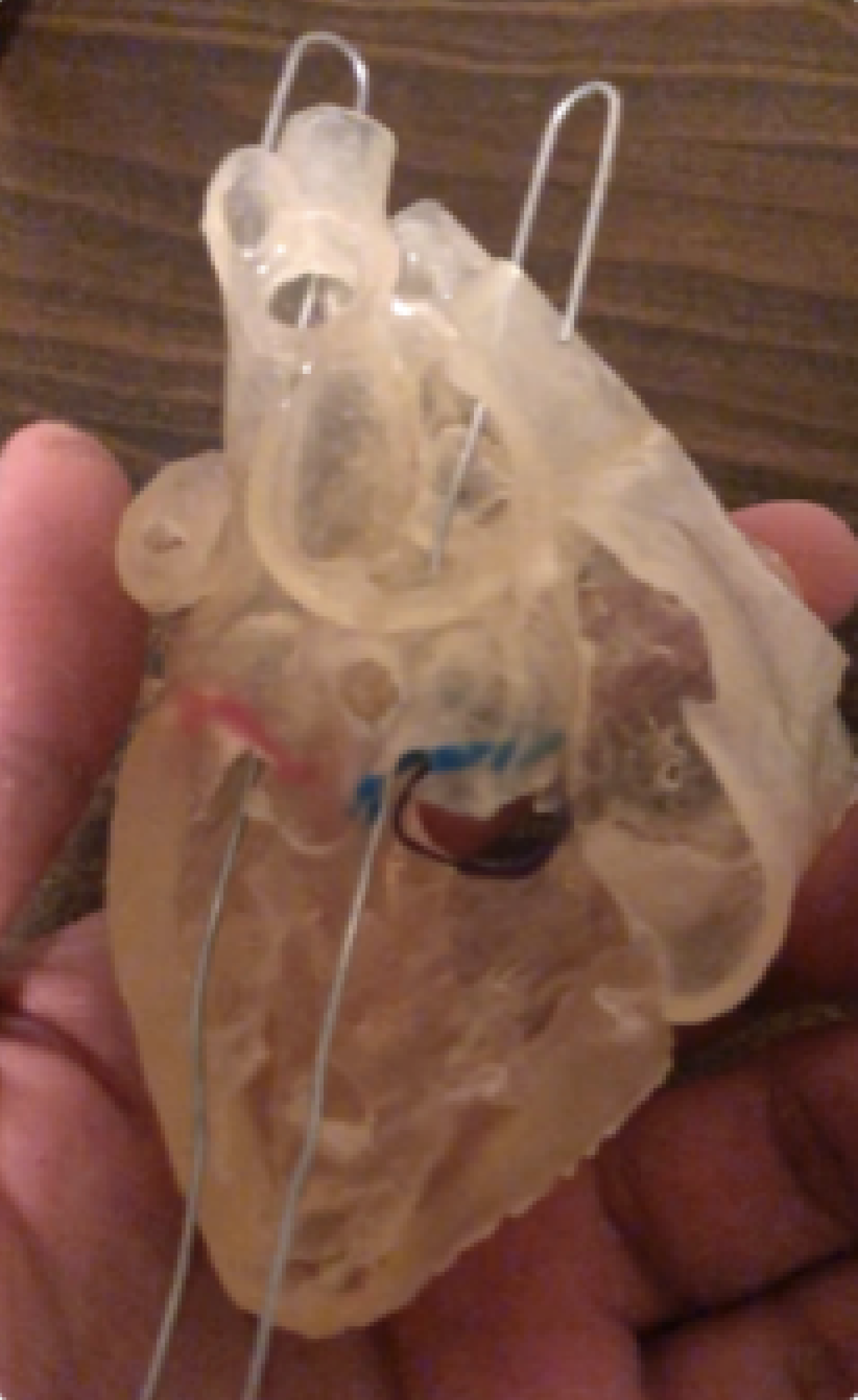}
  \caption{An example of a 3D-printed heart model}
  \label{fig:heart-cutout}
    \end{center}
\end{wrapfigure}

Treating \textit{congenital heart diseases} (CHDs) is a meticulous process that requires surgeons to have strong knowledge of the cardiovascular anatomy of the specific patient, calling for surgical planning processes. Typical surgical planning processes for treating CHDs involve obtaining a 3D-printed heart model constructed using a series of \textit{magnetic resonance imaging} (MRI) scans and \textit{computed tomography} (CT) scans to visualize the morphological features of the patient's heart, as shown in Figure \ref{fig:heart-cutout} \cite{riggs_3d-printed_2018}. This allows surgeons to collaboratively conduct inspections on the heart model and plan a safe surgery strategy \cite{riggs_3d-printed_2018}.

However, producing these models can be time-consuming, resource-intensive, and irreversible, prompting the exploration of alternate technology \cite{Sun2019PersonalizedTP, Gr2018AdaptiveA, Kappanayil2017ThreedimensionalprintedCP, yoo_3d_2021, sun_3d_2022}. Recent research has proposed mixed reality as a promising avenue in medicine. 
Schott, et al. developed a multi-user interface for liver anatomy education that support virtual reality (VR) and augmented reality (AR) environments \cite{schott_vrar_2021}. 
Multiple users could put on head-mounted displays (HMDs) and join a shared environment to interact with various informative elements 
to learn about liver anatomy. 
They evaluated their prototype with medicinal experts in a VR lab and concluded that it achieved good usability. 
The participants preferred the AR environment over VR since it preserved both the surroundings and other participants in view. 
To work towards AR, researchers developed a tool for the HoloLens 2 to render \textit{Digital Imaging and Communications in Medicine} (DICOM) images for surgical planning in an AR environment and allowed users to interact with them through hand-gestures \cite{zhang_directx-based_2022}. 
However, multiple studies have highlighted that HMDs such as the HoloLens 2 are typically uncomfortable to wear for extended periods of time and they require users to learn unfamiliar gesture interactions that can be hard to learn to use 
\cite{dass_augmenting_2018, hirzle_critical_2021}. 

Mobile AR has the potential to address the above challenges by providing a more intuitive and familiar interaction environment with simple gestures that users are already accustomed to \cite{dass_augmenting_2018}.  
Leo, et al. introduced CardiacAR, a mobile AR tool
for cardiovascular surgical planning \cite{Leo21}. 
Yang, et al. evaluated its usability with cardiologists and surgeons, who praised the ease-of-use of mobile AR gestures and highlighted the tool's portability \cite{yang_evaluating_2022}.
However, the cardiologists and surgeons 
noted the absence of collaborative surgical planning support in CardiacAR, which was a limitation as it functioned as a single-user application.
They emphasized the importance of collaboration in facilitating group communication, a key aspect of traditional surgical planning involving 3D heart models
\cite{yang_evaluating_2022}.

Given the advantages of mobile AR, there emerges a need to explore the design and development of a collaborative surgical planning tool harnessing mobile AR benefits for ease-of-use and portability. 
Until now, such exploration is an \textbf{uncharted} territory. 
To address this research gap, our ongoing work makes the following major contributions:

\begin{enumerate}
\item \textbf{Enabling collaborative surgical planning via multi-user augmented reality tools}. In collaboration with surgeons and cardiologists at a large hospital, we are developing \textbf{\tool{}}, an iOS AR application that aims to enable the collaborative cardiovascular surgical planning process by allowing multiple surgeons and cardiologists to join a shared AR environment and inspect 3D models of patients' hearts along with others. 
Our work presents a novel design that combines Apple's peer-to-peer network framework and its state-of-the-art mobile AR framework to develop \tool{}. 
To the best of our knowledge, \tool{} is the \textit{first-of-its-kind} surgical planning tool that utilizes multi-user mobile AR in iOS.
Figure \ref{fig:teaser}A illustrates how multiple surgeons and cardiologists would collaborate, each with their own device to collaboratively inspect the heart from multiple perspectives, through real-time transformations such as scaling and rotation (Figure \ref{fig:teaser}B).
Our ongoing work further extends the feature set to support virtual annotation, allowing surgeons to label important regions on the model. \\

\item \textbf{Streamlining iOS mobile deployment to facilitate broader technology access}. Since \tool{} is built natively on iOS, it can leverage not just the portability and ubiquity of iOS devices, but also their ability to recognize a wide range of gestures and interactions that are intuitive and easy-to-use \cite{dass_augmenting_2018}. 
We are working to enhance \tool{} with planned usability studies and long-term evaluation to gather feedback about relevant features that can further aid with surgical planning. 
By developing within the iOS ecosystem, we  also take advantage of its TestFlight service for rapid deployment and iterative testing of our tool. By making our tool available on the iOS App Store, we will be able to better distribute the application to more surgeons and health care providers easily.

\end{enumerate}

\section{System Design and Implementation}

\begin{figure*}
  \centering\includegraphics[width=\textwidth]{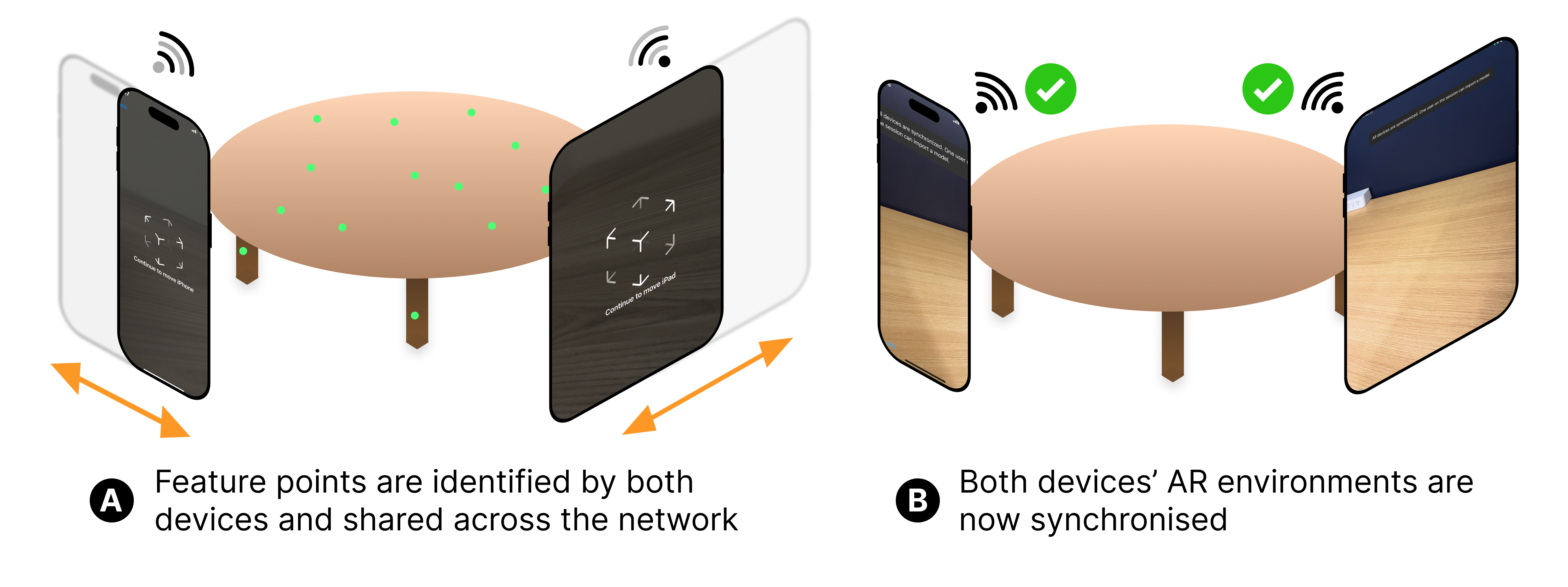}
  \par
  \caption{(A) To initiate a shared session, two surgeons can use their devices to scan their physical environment to detect feature points in their surroundings (shown as green dots), which are notable environmental elements that a device can consistently track across frames, such as the corner of a wooden table. (B) Once the feature points are shared over the network, a pop-up notification is used to inform the user that the AR environments are aligned and the collaborative experience is established.}
          \label{fig:camera-sync}

  \vspace*{\floatsep}
\end{figure*}

\tool{} is developed in Swift and uses ARKit and RealityKit, native iOS frameworks that leverage the device sensors to render 3D objects in the real world. It uses SwiftUI, Apple's newest framework for building user interfaces within the native Apple operating ecosystem. 
Although older frameworks such as UIKit have more documentation and support within iOS AR development, using SwiftUI enables our application to leverage the newest tools offered by Apple.  
To the best of our knowledge, there has been no prior exploration on designing collaborative, multi-user surgical planning in mobile AR. 
\tool{} is the first application that enables this. 
To facilitate its novel application, \tool{} offers the following functionalities:

\subsection{Creating a shared physical space}

When a user opens the application and navigates to the AR View, the device will first check for any existing shared session nearby. If not, it will create a new session that other users can join. Once multiple users in the same physical space open the application, they will all be connected to the same session. This is achieved using Apple's Bonjour services, which is a zero-configuration network protocol that connects external devices to a device's network without needing to assign the external device a specific IP address and entering the address on the device \cite{BonjourConcepts}. These services are implemented in Apple's Multipeer Connectivity framework, allowing the user to detect devices on the network, connect them to the same session, and transfer heart model files and model manipulation data between devices \cite{MultipeerConnectivity}. To create the shared session, the framework utilizes infrastructure Wi-Fi if the devices are on the same network, and peer-to-peer Wi-Fi or Bluetooth if they are not \cite{MultipeerConnectivity}. 

After entering the AR View and joining a session, the users scan the physical space around them using their device camera. 
This allows for key feature points in the environment to be identified \cite{noauthor_arsessioncollaborationdata_nodate}, as shown in Figure \ref{fig:camera-sync}A. 
Feature points are notable environmental elements that a device can consistently track across frames, such as the corner of a wooden table. These points are significant as they offer reliable markers for the device's spatial tracking algorithms.
This information is shared across devices and used to calibrate and recognize the location of each device in the physical space. Once calibrated, the devices display a pop-up notification to the user indicating that the collaborative experience has been established.

\subsection{Interacting with a model in a multi-user environment}

\begin{figure*}
    \begin{center}
        \includegraphics[width=\textwidth]{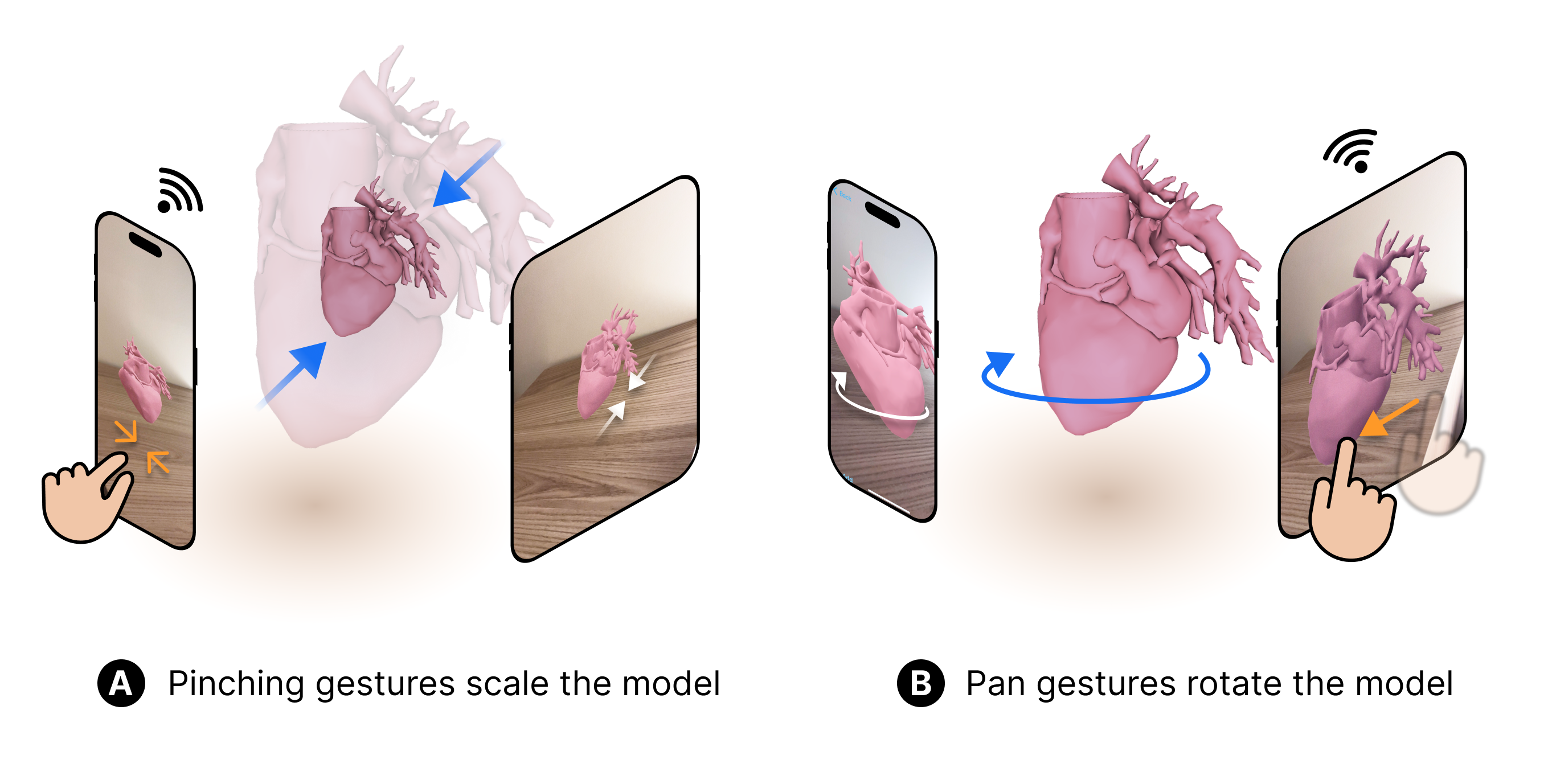}
        \caption{Gestures are used to manipulate the model. The orange arrows represent the gesture performed on one device, the white arrows represent resultant movement of the heart on other devices in the session, and the blue arrows represent actual manipulation of the model. (a) The user can perform a pinch gesture to scale the heart model up or down. (b) The user can pan across the screen with their finger to trigger a rotation of the heart in the direction of the pan.}
        \label{fig:interactions}
    \end{center}
\end{figure*}

When a surgeon or cardiologist imports a model into \tool{}, it is projected into the AR environment through \textit{raycasting}.
To accomplish this, the device casts a ray from the center of the screen and uses it to detect horizontal planes in front of the device. It then proceeds to anchor the model to the plane and triggers this model to render on the other devices as well.

Once the model renders on-screen, \tool{} enables surgeons to interact with the model by rotating and scaling it. 
A pan gesture using one finger is used to rotate the heart, while a pinch gesture using two fingers allows the user to scale the heart up and down. 
Being a collaborative, multi-user tool, \tool{} supports real-time syncronization of the model across all connected devices. 

To achieve this, one approach was to continuously transmit across the network the heart's orientation in the form of a matrix throughout the duration of the shared session. 
However, this approach would cause a lot of memory overhead on the network and would increase latency, making it laggy. 
Instead, we decided to transmit the transformation matrix with every pan and pinch gesture, allowing receiving devices to apply that transformation.

There are multiple steps to compute the transformation matrix from the gestures. 
First, the x and y components of the pan translation vector are utilized to determine the direction of the pan gesture. 
These components are then converted into angular measurements, forming the basis for quaternion calculations. 
Distinctly, this involves generating individual quaternions for each axis. 
The quaternion corresponding to the x-axis is derived based on the pan's x-component, and similarly, the y-axis quaternion is based on the y-component. 
These rotational quaternions are calculated in reference to the camera's current orientation, ensuring alignment with the current camera's perspective.

This procedure results in two separate quaternion rotations, which are then multiplied together. The combined quaternion effectively integrates rotations along both the x and y axes, and is then applied to the orientation matrix of the heart model, following a normalization process to maintain proper rotational orientation. 

However, Multipeer Connectivity only supports sharing across the network if the data can be decoded and encoded \cite{MultipeerConnectivity}, which a quaternion cannot. As a fix, we created a wrapper class with 4 properties, holding the \verb|x, y, z| and \verb|w| properties of the quaternion as floats. This way, we could share this across the network and have the receiving devices convert this to a quaternion and apply the correct transformation.

Additionally, as part of our ongoing work, we are working on an annotation feature that enables surgeons and cardiologists to further interact with the model by labelling certain regions for future discussion. The surgeons and cardiologists in Yang, et al.'s evaluation mentioned that the annotation feature in CardiacAR would be ``very helpful in practical scenarios for labelling and demarcating important regions"\cite{yang_evaluating_2022}. This feature would involve a technique called \textit{hit-testing}, wherein the device casts an invisible ray from the center of the screen and uses it to find the first hit on the heart model.

\subsection{Selecting and sharing a custom heart model}

\begin{wrapfigure}{R}{0.3\textwidth}
    \begin{center}
        \includegraphics[width=0.33\textwidth]{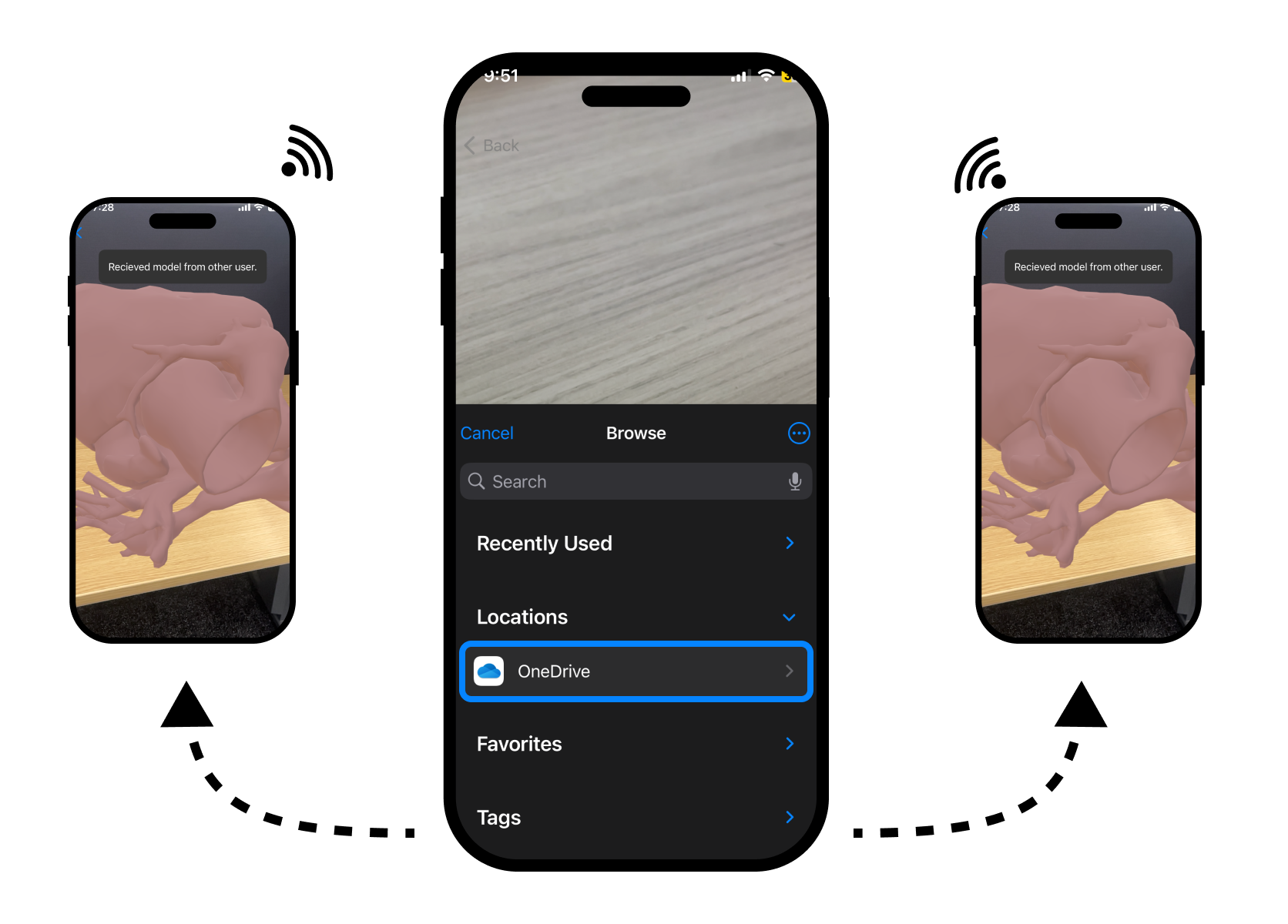}
        \caption{The surgeons and cardiologists can select a 3D model from their HIPAA-compliant OneDrive storage options and the model will be shared to other devices on the network so that they can render it.}
        \label{fig:model-upload}
    \end{center}
\end{wrapfigure}

In Yang, et al.'s usability evaluation, the ``custom model import" feature received an average Likert rating of 4.5 across surgeons and cardiologists due to its facility in supporting patient-specific diagnosis. As part of the feedback, they noted that adding the ability to directly import models from HIPAA\footnote{\textit{Health Insurance Portability and Accountability Act}}-compliant cloud solutions like OneDrive would promote the collaboration process between medical practitioners. For \tool{}, we implement the custom model importing feature and extend the collaboration features by adding the ability to directly import models through OneDrive and sharing imported heart model files to other peers on the shared session using the Multipeer Connectivity Framework.        

Upon establishing a shared session with other users, the host can press the ``Add" button on the screen to open their device file browser and select a 3D file from their local storage. If the medical practitioner is signed into OneDrive, they can select OneDrive from the ``Browse" panel and access their files on the in-app file browser, as shown in Figure \ref{fig:model-upload}. This is achieved using Apple's FileManager interface. 

Once the user selects and confirms the file to display, it must be shared to all the peers on the network so that they are also able to view it. A byte stream is opened and the file is transferred as a sequence of bytes across the network. On the receiving devices, these bytes are then written into memory so that the application can access them. This facilitates the collaboration by requiring only one user to have the model rather than needing every user to individually load the model.

\section{Usage scenario: surgeons and cardiologists planning cardiovascular surgery}
We will present a usage scenario that demonstrates how \tool{} can enable collaboration when multiple surgeons are planning for cardiovascular surgery.

Alex is a surgeon who needs to conduct an operation on a patient with a form of congenital heart disease. This is a time-sensitive matter and Alex would like to get opinions from some senior surgeons and cardiologists. Normally, he would follow the traditional method of 3D-printing the physical model and inspecting it as shown in Figure \ref{fig:heart-cutout}, which could take hours to days as the high-resolution model requires a specialized 3D printer that the hospital must obtain from an external 3D printing company.
Since time is constrained, Alex chooses to use \tool{} to collaboratively interact with the heart model with other surgeons. His cardiologist conducts a CT scan, uses those images to construct a 3D model of the patient's heart, and sends it to Alex via HIPAA-compliant OneDrive file sharing. 

Similar to the logistics of how typical surgical planning sessions would be performed when using a 3D-printed heart, Alex convenes in a room with Lisa, a senior surgeon,  and Jonathan, a cardiologist, to inspect the heart model and discuss surgical plans.
All of them open up \tool{} on their own iOS devices. Alex and Lisa are using their iPhones because they are frequently on the move, while Jonathan uses an iPad. 
They use their devices' cameras to scan the surroundings for key feature points, as shown in Figure \ref{fig:camera-sync}A, to align their AR environments. 
Once they see the notification indicating that their devices are synchronised, Alex imports the heart model from OneDrive by navigating through the in-app file browser, as shown in Figure \ref{fig:model-upload}. 
This model is shared across the network and it is rendered on all 3 devices. Alex uses a pinch gesture with two fingers to scale up the heart model, similar to the demonstration in Figure \ref{fig:interactions}A, so that the other surgeons can see it more clearly. Since they are not restricted by the weight and limited movements of a head-mounted display, they can freely move their devices around the heart to view it from different perspectives. To save time, they each move their devices through different tracts of the heart to observe and discuss any abnormalities they see. To view the pericardium of the heart model, Alex pans across the screen to rotate the heart model such that it is visible for everyone, as shown in Figure \ref{fig:interactions}B. The session helps Alex get some clarity on certain structural questions he initially had about the heart model.

Using \tool{} for collaborative multi-user surgical planning gave Alex the idea that it would be very useful in an educational setting. 
Alex recalled that when he was a surgical resident, he had few opportunities to participate in surgical planning for congenital heart disease due to the limited number of surgeries and the resources needed to reproduce the physical 3D heart models. With \tool{}, surgical residents could more easily inspect a variety of heart models without having to go through the 3D printing process and gain more exposure to the surgical planning process. 

\section{Conclusion and Ongoing Work}
We have presented our ongoing research \tool{} that focuses on designing and developing the enabling technological capabilities to support multi-user collaborative surgical planning in a mobile AR environment. \tool{} offers various features such as multi-user AR sessions, custom model import, and model manipulation that allow it to be used in real-world cardiovascular surgical planning scenarios. We plan to build on this foundation to integrate more features and to evaluate them with surgeons and cardiologists.

\subsection{Supporting Collaborative Model Slicing}
In our ongoing research, we are expanding its features, supporting an additional interaction: omni-directional slicing. This feature will allow surgeons and cardiologists to view different cross-sections of the heart model as they do in the current surgical planning process involving physical models \cite{riggs_3d-printed_2018}. Additionally, similar functionality has been previously implemented for single-person use, and was highly rated by surgeons and cardiologists \cite{yang_evaluating_2022}. To enable this, we are adding a dedicated ``slicing mode'' where users can manipulate a 3D slicing plane to view the cross-sections. Since model slicing is not a native feature supported in RealityKit or ARKit, we are developing our own approach in RealityKit for multi-user omni-directional model slicing using Apple's Metal surface shader to conditionally render polygons on one side of the slicing plane. 

\subsection{Evaluation with Surgeons and Cardiologists}
To evaluate the real-world impact of the novel collaborative multi-user mobile AR technology, we plan to conduct a user study. This study will occur in two phases.

In the first phase, we will evaluate \tool{} with the surgeons and cardiologists in our team, similar to how Yang et al. \cite{yang_evaluating_2022} evaluated the single-user CardiacAR. The participants will use \tool{} to perform a series of tasks as a group (e.g., importing a model, scaling it, annotating it, performing omni-directional slicing), and then rate aspects such as usability and usefulness on a Likert scale. 
In the second phase, to evaluate to the long-term real-world impact of \tool{}, our surgeon and cardiologist collaborators will experiment with conducting a surgical planning session using \tool{} and compare the experience with their current approach of using physical 3D heart model. The application will be uploaded to Apple's TestFlight service, allowing the participants to install it on their own device. After both sessions conclude, we will ask them questions to understand how \tool{} compares to the existing approach.
We will use the feedback from this two-part user study to further improve the features and user interface of \tool{}. 

\subsection{Deployment on iOS App Store}
We will be deploying \tool{} on the iOS App Store to make it available and free-to-use for surgeons and cardiologists who have an iOS device. This will broaden testing and allow us to collect additional feedback from surgeons and cardiologists in different parts of the world.

\section{Acknowledgements}
The heart model in the \tool{} logo in Figure \ref{fig:teaser} is adapted from ``Medical Heart'' by Team Iconify under a Creative Commons Attribution License 3.0.

\bibliographystyle{ACM-Reference-Format}
\bibliography{Collaborative AR}

\end{document}